\documentstyle[aps,prc,twocolumn,psfig]{revtex}

\begin{document}




\newcommand{\ep}{equivalence principle~}

\newcommand{\Ch}{Chandrasekhar~}

\newcommand{\Chp}{Chandrasekhar}

\newcommand{\Sc}{Schwarzschild~}

\newcommand{\Scp}{Schwarzschild}

\newcommand{\Sw}{Schwarzschild~}

\newcommand{\Swp}{Schwarzschild}

\newcommand{\Sch}{Schr{\"{o}}dinger~}

\newcommand{\Schp}{Schr{\"{o}}dinger}

\newcommand{\OVp}{Oppenheimer--Volkoff}

\newcommand{\OV}{Oppenheimer--Volkoff~}

\newcommand{\GR}{General Relativity~}

\newcommand{\GT}{General Theory of Relativity~}

\newcommand{\GRp}{General Relativity}

\newcommand{\GTp}{General Theory of Relativity}

\newcommand{\STp}{Special Theory of Relativity}

\newcommand{\ST}{Special Theory of Relativity~}

\newcommand{\Lt}{Lorentz transformation~}

\newcommand{\Ltp}{Lorentz transformation}

\newcommand{\rel}{relativistic~}

\newcommand{\relp}{relativistic}

\newcommand{\msun}{M_{\odot}}

\newcommand{\eos}{equation of state~}

\newcommand{\eoss}{equations of state~}

\newcommand{\eossp}{equations of state}

\newcommand{\eosp}{equation of state}

\newcommand{\Eos}{Equation of state}

\newcommand{\Eosp}{Equation of state}

\newcommand{\beqn}{\begin{eqnarray}}

\newcommand{\eeqn}{\end{eqnarray}}

\newcommand{\nonum}{\nonumber \\}

\newcommand{\walecka}{$\sigma,\omega,\rho$~}

\newcommand{\waleckap}{$\sigma,\omega,\rho$}

\newcommand{\bbar}[1] {\mbox{$\overline{#1}$}} 

\newcommand{\courtesy}{~Reprinted with permission of Springer--Verlag 

New York; copyright 1997}

\newcommand{\oo}{{\"{o}}}

\newcommand{\au}{{\"{a}}}


\newcommand{\approxlt} {\mbox {$\stackrel{{\textstyle<}} {_\sim}$}}

\newcommand{\approxgt} {\mbox {$\stackrel{{\textstyle>}} {_\sim}$}}

\newcommand{\mearth} {\mbox {$M_\oplus$}}

\newcommand{\rearth} {\mbox {$R_\oplus$}}

\newcommand{\eo} {\mbox{$\epsilon_0$}}


\newcommand{\bfalpha} {\mbox{\mbox{\boldmath$\alpha$}}}

\newcommand{\bfgamma} {\mbox{\mbox{\boldmath$\gamma$}}}

\newcommand{\bfrho} {\mbox{\mbox{\boldmath$\rho$}}}

\newcommand{\bfsigma} {\mbox{\mbox{\boldmath$\sigma$}}}

\newcommand{\bftau} {\mbox{\mbox{\boldmath$\tau$}}}

\newcommand{\bfLambda} {\mbox{\mbox{\boldmath$\Lambda$}}}

\newcommand{\bfpi} {\mbox{\mbox{\boldmath$\pi$}}}

\newcommand{\bfomega} {\mbox{\mbox{\boldmath$\omega$}}}


\newcommand{\bp} {\mbox{\mbox{\boldmath$p$}}}

\newcommand{\br} {\mbox{\mbox{\boldmath$r$}}}

\newcommand{\bx} {\mbox{\mbox{\boldmath$x$}}}

\newcommand{\bv} {\mbox{\mbox{\boldmath$v$}}}

\newcommand{\bu} {\mbox{\mbox{\boldmath$u$}}}

\newcommand{\bk} {\mbox{\mbox{\boldmath$k$}}}

\newcommand{\bA} {\mbox{\mbox{\boldmath$A$}}}

\newcommand{\bB} {\mbox{\mbox{\boldmath$B$}}}

\newcommand{\bF} {\mbox{\mbox{\boldmath$F$}}}

\newcommand{\bI} {\mbox{\mbox{\boldmath$I$}}}

\newcommand{\bJ} {\mbox{\mbox{\boldmath$J$}}}

\newcommand{\bK} {\mbox{\mbox{\boldmath$K$}}}

\newcommand{\bP} {\mbox{\mbox{\boldmath$P$}}}

\newcommand{\bS} {\mbox{\mbox{\boldmath$S$}}}

\newcommand{\bdel} {\mbox{\mbox{\boldmath$\bigtriangledown$}}}


\newcommand{\eps} {\mbox {$\epsilon$}}

\newcommand{\gpercm} {\mbox {${\rm g}/{\rm cm}^{3}$}}

\newcommand{\rhon} {\mbox {$\rho_{0}$}}

\newcommand{\rhoc} {\mbox {$\rho_{c}$}}

\newcommand{\fmm} {\mbox {${\rm fm}^{-3}$}}

\newcommand{\bag} {\mbox {$B^{1/4}$}}


\newcommand{\fraca} {\mbox {$ \frac{1}{2}       $}}

\newcommand{\fracb} {\mbox {$ \frac{3}{2}       $}}

\newcommand{\fracc} {\mbox {$ \frac{1}{4}       $}}

\newcommand{\fraccc} {\mbox {$ \frac{1}{3}       $}}

\newcommand{\xn} {\mbox {$ x_{N}     $}}
\newcommand{\un} {\mbox {$ u_{N}    $}}
\newcommand{\vlambda} {\mbox {$ \frac{1}{4} \lambda (\sigma^{2} -
\sigma_{0}^{2} )^{2}      $}}
\newcommand{\kf} {\mbox {$ k_{F}       $}}
\newcommand{\stressenergy} {\mbox {$
\calT_{\mu \nu} = -g_{\mu \nu} \calL  + \sum_{\phi} \frac{\partial \calL}
{\partial (\partial^{\mu} \phi)} \partial_{\nu} \phi $}}
\newcommand{\mstar}[1] {\mbox {$ m^{\star}_{#1}      $}}
\newcommand{\ef}[2] {\mbox {$ (\kf^{2} + \mstar{#2}^{2})^{#1 /2}  $}}
\newcommand{\ek}[2] {\mbox {$ (k^{2} + \mstar{#2}^{2} )^{#1 /2}  $}}
\newcommand{\msigma} {\mbox {$ m_{\sigma}      $}}
\newcommand{\momega} {\mbox {$ m_{\omega}      $}}
\newcommand{\mrho} {\mbox {$ m_{\rho}      $}}
\newcommand{\mN} {\mbox {$ m_{N}      $}}
\newcommand{\mB} {\mbox {$ m_{B}      $}}
\newcommand{\munu} {\mbox {$ \mu \nu       $}}
\newcommand{\calT} {\mbox {$ {\cal T}       $}}
\newcommand{\calH} {\mbox {$ {\cal H}       $}}
\newcommand{\calL} {\mbox {$ {\cal L}       $}}
\newcommand{\calD} {\mbox {$ {\cal D}       $}}
\newcommand{\Lint}[1] {\mbox {$ \bar{\psi}_{#1} [ g_{\omega }
\gamma_{\mu} \omega^{\mu} + \frac{1}{2} g_{\rho }  \gamma_{\mu}
 \bftau\! \cdot\! \bfrho^{\mu} \cdots  ] \psi_{#1}  $}}
 \newcommand{\LLint}[1] {\mbox {$ \bar{\psi}_{#1} [ - g_{\sigma}
 \sigma + g_{\omega }
 \gamma_{\mu} \omega^{\mu} + \frac{1}{2} g_{\rho }  \gamma_{\mu}
  \bftau\! \cdot\! \bfrho^{\mu} \cdots  ] \psi_{#1}  $}}
  \newcommand{\Lsigma}[1] {\mbox {$
  \bar{\psi }_{#1} [i \gamma_{\mu} \partial^{\mu} -g( \sigma
  +i \gamma_{5}\bftau\! \cdot\! \bfpi) ] \psi_{#1}  $} }
  \newcommand{\Lmeson} {\mbox {$
  \frac{1}{2} (\partial_{\mu} \sigma \partial^{\mu} \sigma
    + \partial_{\mu}\bfpi\! \cdot\! \partial^{\mu}\bfpi)
      - \frac{1}{4} \lambda (\sigma^{2}+\bfpi\! \cdot \!
	\bfpi- \sigma_{0}^{2})^2 $} }

\newcommand{\EM}{\sqrt{k^2+m^{2}_{M}}}
\newcommand{\E}{\sqrt{k^2+m^{\star 2}}}
\newcommand{\EEE}{\bigl(k^2+m^{\star 2}\bigr)^{3/2}}
\newcommand{\EE}{\mbox{$ \sqrt{ k^{2}+(m-g_{\sigma}\sigma)^{2} } $}}
\newcommand{\ms}{\mbox{$ m-g_{\sigma}\sigma  $}}
\newcommand{\Ethree}{(k^2+m^{\star 2})^{3/2}}
\newcommand{\LLL}[1]{ {\cal L}_{#1}^{0} }
\newcommand{\gs}{ g_\sigma }
\newcommand{\gth}{ g_{3\sigma} }
\newcommand{\gfo}{ g_{4\sigma} }
\newcommand{\gv}{ g_\omega }
\newcommand{\gr}{ g_\rho }
\newcommand{\UU}{\mbox{$
 \frac{1}{3} b m (g_{\sigma} \sigma)^{3}
 + \frac{1}{4} c(g_{\sigma}\sigma)^{4}  $}}
 \newcommand{\U}{\mbox{$
  \frac{1}{3} b m_{n} (g_{\sigma} \sigma)^{3}
  + \frac{1}{4} c(g_{\sigma}\sigma)^{4}  $}}
  \newcommand{\Um}{\mbox{$
  -\; \frac{1}{3} b m_{n} (g_{\sigma} \sigma)^{3}
  - \frac{1}{4} c(g_{\sigma}\sigma)^{4}  $}}
  \newcommand{\UUm}{\mbox{$
  -\; \frac{1}{3} b m (g_{\sigma} \sigma)^{3}
  - \frac{1}{4} c(g_{\sigma}\sigma)^{4}  $}}
  \newcommand{\Upm}{\mbox{$ -\;b m_n (g_{\sigma} \sigma)^2
  -c (g_{\sigma} \sigma)^3  $}}
  \newcommand{\M}{\mbox{$ \sqrt{ k^{2}+(m_{B}-g_{\sigma B}\sigma)^{2} } $}}
  \newcommand{\Mstar}[1]{\mbox{$ \sqrt{ k^{2}+\mstar{#1}^2 } $}}
  \newcommand{\m}{\mbox{$ \sqrt{ k^{2}+m_{\lambda}^{2} }  $}}
  \newcommand{\ke}[1]{\mbox{$ \frac{1}{2} m_{#1}^{2} #1^{2}  $}}
  \newcommand{\keo}{\mbox{$\frac{1}{2} m_{\omega}^2 \omega_{0}^{2} $}}
  \newcommand{\kerho}{\mbox{$\frac{1}{2} m_{\rho}^2 \rho_{03}^{2} $}}
  \newcommand{\meff}{m_{B}-g_{\sigma B}\sigma}
  \newcommand{\coup}[1]{\mbox{$ (g_{#1}/m_{#1})^{2} $}}
  \newcommand{\lsigma}{\mbox{$\frac{1}{2}(\partial_{\mu} \sigma
  \partial^{\mu} \sigma - m_{\sigma}^{2} \sigma^{2})$  }}
  \newcommand{\lsigmaa}{\mbox{$\frac{1}{2}(\partial_{\mu} \sigma_{1}
  \partial^{\mu} \sigma_{1} - m_{\sigma}^{2} \sigma_{1}^{2})$  }}
  \newcommand{\lsigmab}{\mbox{$\frac{1}{2}(\partial_{\mu} \sigma_{2}
  \partial^{\mu} \sigma_{2} - m_{\sigma}^{2} \sigma_{2}^{2})$  }}
  \newcommand{\lsigmastar}{\mbox{$(\partial_{\mu} \sigma^\ast
  \partial^{\mu} \sigma - m_{\sigma}^{2} \sigma^\ast \sigma)$  }}
  \newcommand{\lomega}{\mbox{$ - \: \frac{1}{4} \omega_{\mu \nu}
  \omega^{\mu \nu} +\frac{1}{2} m_{\omega}^{2} \omega_{\mu} \omega^{\mu} $}}
  \newcommand{\lrho}{\mbox{$
  - \: \frac{1}{4}\bfrho_{\mu \nu}\! \cdot\! \bfrho^{\mu \nu}
  + \frac{1}{2} m_{\rho}^{2}\bfrho_{\mu}\! \cdot\! \bfrho^{\mu} $}}
  \newcommand{\lomunu}{\mbox{$ \omega_{\mu \nu} =
  - \: \frac{1}{4}\bfrho_{\mu \nu}\! \cdot\! \bfrho^{\mu \nu}
  + \frac{1}{2} m_{\rho}^{2}\bfrho_{\mu}\! \cdot\! \bfrho^{\mu} $}}
  \newcommand{\omunu}{\mbox{$ \omega_{\mu \nu} =
  \partial_{\mu} \omega_{\nu}
  - \partial_{\nu} \omega_{\mu} $}}
  \newcommand{\nbaryon}[1] { \mbox{$
      \bigl(  \exp [(\epsilon_{B}(k) {#1})/T] +1  \bigr) ^{-1} $ } }
      \newcommand{\nmeson}[2] {\mbox{$
	  \bigl(  \exp [(\omega_{{#2}} (k) {#1})/T] -1 \bigr) ^{-1} $ } }
	  \newcommand{\D} {\mbox{$ {\cal D}(k,\omega) $} }
	  \newcommand{\Dm} {\mbox{$ {\cal D}^{-1}(k,\omega) $} }
	  \newcommand{\slpartial} {\mbox{$ \not\!\partial $} }
	  \newcommand{\slgamma} {\mbox{$  \not\!\gamma $} }
	  \newcommand{\sldel} {\mbox{$  \not\!\!\nabla $}}
	  \newcommand{\slp} {\mbox{$  \not\!\! p  $}}
	  \newcommand{\slA} {\mbox{$  \not\!\! A  $}}
	  \newcommand{\slK} {\mbox{$  \not\!\! K  $}}


\newcommand{\tit} {Neutron Star Constraints on the H Dibaryon}

\newcommand{\auth} {Norman K. Glendenning}

\newcommand{\lbl}{LBNL-46879}

\newcommand{\dateofdoc}{\today}

\newcommand{\adr} {Nuclear Science Division \& 
Institute for Nuclear and Particle Astrophysics,
  Lawrence Berkeley  National Laboratory,
   MS: 70-319 \\ Berkeley, California 94720}


\newcommand{\doe}
{This work was supported by the
Director, Office of Energy Research,
Office of High Energy
and Nuclear Physics,
Division of Nuclear Physics,
of the U.S. Department of Energy under Contract
DE-AC03-76SF00098.}

\draft
\title{On the uncertainty of hyperon couplings and the electrochemical
potential in neutron star matter}
\author{Norman K. Glendenning}
\address{Nuclear Science Division and
Institute for Nuclear \& Particle Astrophysics,
 Lawrence Berkeley National Laboratory, MS: 70-319, Berkeley,
California 94720}
\date{\today}
\maketitle

\begin{abstract}
Uncertainty of the hyperon couplings, in particular, that of the
$\Sigma^-$, in dense matter raises the question
of the behavior of the electrochemical potential
in neutron star matter, which is crucial to the
possible presence of the kaon condensed phase. 
We show that regardless of this uncertainty, the $\Lambda$ hyperon,
whose coupling can be constrained by its binding in nuclear matter and other
observations,  also aided by the $\Xi^-$,
introduce a saturation of the electrochemical potential just as the
$\Sigma^-$ would otherwise do, which tends to
mitigate against kaon condensation.
The maximum possible mass of neutron stars appears to be
$\sim 1.5 \msun$ independent of the uncertainties.
\end{abstract}


\section{INTRODUCTION}

Hyperons, the strange members of the baryon octet, are likely to exist in
high density matter and in particular its charge neutral form, often
referred to as neutron star matter. The Pauli principle practically
assures that their presence will lower the Fermi energy of baryon species
and hence the total energy at given baryon number. However, aside from this
general argument, the coupling constants of hyperons also influences the
extent of their participation. The $\Lambda$ hyperon is a  partial 
exception to
this uncertainty  \cite{glen91:c}. 
Its couplings can be at least constrained by the
experimentally extrapolated value of its binding in nuclear matter
\cite{dover88:a}, by the results of an analysis of
hypernuclear levels \cite{rufa90:a}, and
the requirement that theory can account for neutron stars of mass as 
great as $1.5 \msun$.

The important neutron star properties that hyperons effect are the limiting
neutron star mass and the possibility of kaon condensation.
As compared to models populated only by nucleons and
leptons, hyperons reduce the maximum mass by as much as $3/4 \msun$.
The reason their presence strongly effects the possibility of kaon condensation
is as follows
\cite{glen85:b}: The effective mass of kaons
in nuclear matter is reduced from its vacuum mass by an attractive interaction
with the nuclear medium
\cite{kaplan86:a}. If the ${\rm K}^-$ effective mass sinks to a
value of the electron chemical potential as the density of matter increases, 
the ${\rm K}^-$ can thereafter replace the electron as the charge neutralizing
agent in neutron star matter. The kaons can all condense in the zero momentum
state, whereas electrons have to occupy ever higher momentum states with
increasing density. {\sl However}, hyperons may saturate the electron 
chemical potential at a relatively low
density either postponing the appearance of a kaon condensate to a high
density, or 
preempting it altogether. 
The reason that hyperons can do this is because they carry the conserved 
baryon charge and they occur in all three charge states, $\pm 1 {~\rm and~}
0$. Therefore it may happen that charge neutrality can be achieved 
most energetically favorably  among baryons with little participation
of leptons. (Lepton number is
not conserved because of neutrino loss from the
star.) The foregoing conclusions of Ref.\ \cite{glen85:b} have been
confirmed in subsequent work \cite{schaffner96:a,schaffner96:b,balberg97:a}.

That hyperons can  contribute to the saturation
of the electron chemical potential and thereby
preempt the condensation of kaons depends, at first sight,
on the $\Sigma^-$ since it is the lowest mass baryon of negative
charge and can replace a neutron and electron.  Extrapolated atomic data 
suggest that it may feel repulsion at high density, which would mitigate
against its appearance in dense matter, although this remains inconclusive
\cite{gal94:a}.
Indeed, it has been suggested that the absence of the  $\Sigma^-$
would mitigate the negative effect that hyperons have on 
kaon condensation \cite{brown}.

However, we show in this paper, that even if the  $\Sigma^-$ is totally
absent from dense neutral matter, the $\Lambda$ hyperon 
aided by the $\Xi^-$ also causes the
electron chemical potential to saturate and then decrease with increasing
density. 
The  $\Lambda$ is known to experience an attractive potential
in normal nuclear matter
\cite{dover88:a}
as does the $\Xi^-$ \cite{dover83:a,fukuda,khaustov}.
The $\Lambda$ can replace neutrons at the top of their Fermi sea with
 a reduction  in the  high value of
the 3-component of the isospin of neutron star matter, thus reducing the
assymetry energy, and with no
increase in electron population with increase of density.
The $\Xi^-$ can replace a neutron and electron and also 
enhance the proton population at the expense of the neutron, 
just as the $\Sigma^-$,
and has a low density threshold in the absence of the  $\Sigma^-$.
The net effect  is that hyperons disfavor kaon condensation by terminating the
growth of the electron population and electrochemical 
potential with increasing density, even
if the $\Sigma^-$ interaction were so strongly repulsive that it is absent from
neutron star matter in the density range relevant to those stars.

\section{THEORY}

We describe nuclear matter by the mean field solution
of the covariant
Lagrangian
\cite{glen85:b,books,glen82:a,bodmer77:a,%
serot87:a,kapusta90:a,kapusta91:a} which is a generalization of the
model introduced first by Johnson and Teller \cite{teller55:a}, by
Duerr \cite{duerr56:a} and later by Walecka \cite{walecka74:a}:
\begin{eqnarray}
{\cal L} & = &
\sum_{B} \overline{\psi}_{B} (i\gamma_{\mu} \partial^{\mu} - m_{B}
+g_{\sigma B} \sigma  - g_{\omega B} \gamma_{\mu} \omega^{\mu}
\nonumber \\[0ex]
& &
 - \fraca g_{\rho B} \gamma_{\mu} \bftau \cdot \bfrho^{\mu} )
 {\psi}_{B}
 +\:  \lsigma 
 \nonumber \\[2ex]
 & &
   \lomega
     \lrho \nonumber \\[2ex]
     & &  \:\Um 
     \nonumber  \\[2ex]
      & &
       + \sum_{e^{-},\mu^{-}}
     \overline{\psi}_{\lambda} \bigl(i\gamma_{\mu}
       \partial^{\mu} - m_{\lambda} \bigr) \psi_{\lambda}\,.
	 \label{lagrangian}
	 \end{eqnarray}
	 The advantage of the model as compared with other models of
	 nuclear matter, is that it can be made to agree with five
	 nuclear properties at {\sl saturation} density, the highest
	 density for which we have any empirical knowledge, and it
	 extrapolates causally to all densities.
	 The baryon species, denoted by
	 $B$, are coupled to the scalar, vector and vector-isovector
	 mesons,  $\sigma, \omega, \bfrho$.
	  The masses are denoted by $m$ with an appropriate subscript.
	 The sum on $B$ is over all the charge states of the lowest baryon
	 octet, ($p,n,\Lambda,\Sigma^{+},\Sigma^{-}, \Sigma^{0},
	 \Xi^{-}, \Xi^{0}$) as well as the $\Delta$ quartet
	 and the triply strange baryon, $\Omega^-$. However
	 the latter two 
	 are not populated up to the highest density in neutron
	 stars, nor are any other baryon states save those of the lowest
	 octet for reasons given elsewhere \cite{glen85:b}.
	 The cubic and quartic $\sigma$ terms were first introduced by
	 Boguta and Bodmer so as to bring two additional nuclear
	 matter properties under control \cite{bodmer77:a}.
	 The last term represents the free lepton Lagrangians.
	 How the theory can be solved in the mean field
	 approximation for the ground state of charge neutral
	 matter in general beta equilibrium (neutron star matter)
	 is described fully in
	 Refs. \cite{glen85:b,books}.

The mean values of the non-vanishing meson fields are denoted 
by $\sigma,\omega_0, \rho_{03}$,
in which case the baryon effective masses are given by
$\mstar{B} = m_B - g_{\sigma B} \sigma$ and the baryon eigenvalues
by
\beqn
e_{B}(k) =
g_{\omega B} \omega_{0} + g_{\rho B} \rho_{03} I_{3B} +
\sqrt{k^{2} + \mstar{B}^2}.
\label{eigenvalue}
\eeqn
In the above
equations,
$I_{3 B}$ is the isospin projection of baryon charge state B.

The Fermi momenta for the baryons are the positive real solutions
of
\beqn
e_B(k_B)= \mu_B  \equiv b_B \mu_n-q_B\mu_e \,,
\label{equil}
\eeqn
where $b_B$ and $q_B$ are the baryon and electric charge numbers of the
baryon state B, and $\mu_n$ and $\mu_e$ are independent
chemical potentials for unit baryon number and unit negative electric charge
number
(neutron and electron respectively).
 The
 lepton Fermi momenta are the positive real solutions of,
 \beqn
 \sqrt{k^2_{e}+m^2_{e}}=\mu_e\,,~~~~~~
 \sqrt{k^2_{\mu}+m^2_{\mu}}=\mu_{\mu}=\mu_e.
 \label{lepton}
 \eeqn
These equations  (\ref{equil}) and (\ref{lepton})
ensure chemical equilibrium.

 Charge neutrality is expressed 
 as
 \beqn
 q_H & \equiv &\sum_B (2J_{B}+1) q_B
 k_{B}^{3}/(6 \pi^2 ) -
 \sum_{\lambda} k_{\lambda}^{3}/(3 \pi^2 ) =0
 \label{charge}
 \eeqn
 where the first sum is over the baryons whose Fermi momenta are
   $k_B$
   and the second sum is over the leptons
   e$^{-}$ and $\mu^{-}$.  By simultaneously
   solving the meson  field equations, the condition for
    charge neutrality,
    and the conditions for chemical equilibrium
    (\ref{equil},~\ref{lepton}),
    we get the  solution for the three mean fields, 
    the
	two chemical potentials, the two
    lepton Fermi momenta, the N baryon Fermi momenta  
    (where N is the number of baryon
	charge states populated)
     of
    beta-stable charge-neutral matter called neutron-star-matter
    at the chosen baryon density in the hadronic phase,
    \beqn
    \rho=\sum_B (2J_B +1) k^{3}_{B}/(6 \pi^2)\,.
    \eeqn

 The \eos can be calculated at each baryon density for which the
 solution for the  7+N variables  enumerated above have been found.
 It is:
  \\[1ex]
\begin{eqnarray}
\nopagebreak
\epsilon  & = & \ke{\sigma} + \keo
+ \kerho
\nonumber  \\[1ex]
& &
        + \sum_{B} \frac{2 J_{B} + 1}{2 \pi^{2} }
\int_{0}^{k_{B}} \sqrt{k^2 + \mstar{B}^2}
\, k^{2} \, dk 
\nonumber  \\[0ex]
& &
+\sum_{\lambda} \frac{1}{\pi^{2}} \int_{0}^{k_{\lambda}}
\m
\, k^{2} \,  dk,
           \label{emfab}
\end{eqnarray}
which is the energy density while the pressure  is given  by,
\beqn
p & = &  -\ke{\sigma} + \keo
+ \kerho 
\nonumber  \\[1ex]
& &
        + \frac{1}{3} \sum_{B} \frac{2 J_{B} + 1}{2 \pi^{2} }
\int_{0}^{k_{B}} k^{4} \, dk / \sqrt{k^2+\mstar{B}^2 } 
 \nonumber  \\[0ex]
 & &
 +\frac{1}{3}
         \sum_{\lambda} \frac{1}{\pi^{2}} \int_{0}^{k_{\lambda}}
 k^{4} \, dk /  \m.
           \label{pmfa}
\eeqn
These are the diagonal components of the stress--energy tensor
\beqn
\calT^{\mu\nu}= -g^{\mu\nu} \calL
+\sum_{\phi} \frac{\partial \calL}{\partial (\partial_{\mu} \phi)}
\partial^{\nu} \phi\,.
\eeqn

Five of the  constants of the theory can be
algebraically  determined
by the properties of nuclear matter \cite{books}.
The constants are
the nucleon couplings to the
scalar, vector and vector--isovector mesons, $g_\sigma/m_\sigma,
g_\omega/m_\omega, g_\rho/m_\rho$,
and  the scalar self-interactions defined by $b$ and $c$.
The nuclear properties that define their values
used here are  the binding energy $B/A=-16.3$
MeV, baryon density $\rho=0.153 {~\rm fm^{-3}}$,
symmetry energy coefficient $a_{sym}=32.5$ MeV, compression modulus
$K=240$ MeV and
nucleon effective mass $m^*/m=0.78$.
How these choices are related to empirical data is discussed in Chapter 4,
Section 5
of Ref.\ \cite{books}.

\begin{figure}[htb]
\vspace{-.4in}
\hspace{.18in}\psfig{figure=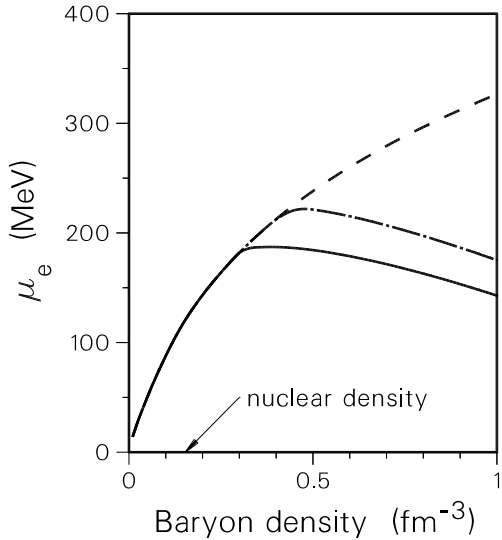,width=2.76in}
\vspace{.2in}
\begin{center}
\parbox[t]{4.6 in} { \caption{ \label{mu3} Electrochemical potential
in neutron star matter as a function of density. Three cases are compared:
(1)
 only nucleons and leptons are present (dashed line), (2) nucleons, hyperons
  and leptons are present (solid line),
   (3) nucleons, leptons and hyperons except the
    $\Sigma^-$ are present (dash-dot line).
    }}
    \end{center}
    \end{figure}

Nuclear matter at normal density does not depend on the hyperon couplings.
Elsewhere we have shown how they can be made consistent with (1) the
data on hypernuclear levels,  (2) the
binding of the $\Lambda$ in nuclear matter
\begin{figure}[htb]
\vspace{-.4in}
\hspace{.18in}\psfig{figure=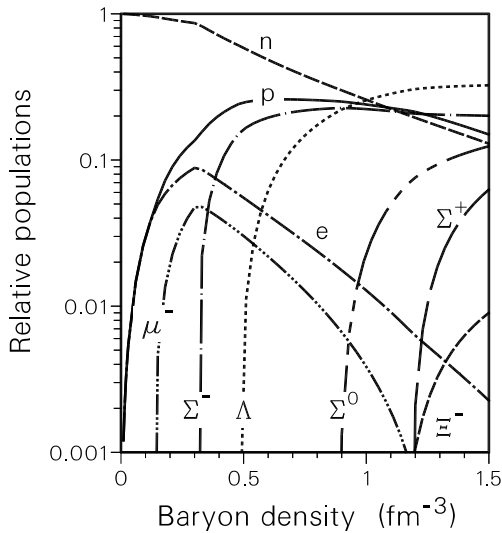,width=2.76in}
\vspace{.2in}
\begin{center}
\parbox[t]{4.6 in} { \caption{ \label{pops_240} Particle populations
in neutron star matter containing nucleons, hyperons and leptons.
}}
\end{center}
\end{figure}
\noindent (which can be determined quite accurately
from  an extrapolation of the hypernuclear levels to large atomic
number $A$),  and (3)  neutron star masses \cite{glen91:c}.
We shall assume that
all hyperons in the octet have the same coupling
as the $\Lambda$.
The couplings are expressed as a ratio to the above mentioned
nucleon couplings,
\beqn
x_\sigma=g_{H\sigma}/g_{\sigma},~~~~
x_\omega=g_{H\omega}/g_{\omega},~~~~
x_\rho=g_{H\rho}/g_{\rho}.
\label{ratio}
\eeqn
The first two are related to the $\Lambda$ binding by a relation derived
in \cite{glen91:c} and the third can be taken equal to the second
by invoking vector dominance.
 Together  the hyperon couplings are limited to the range
  $0.5<x_\sigma <0.7$ 
  \cite{glen91:c} and we take $x_\sigma=0.6$.
The corresponding value of $x_\omega$ is $0.658$.

 \begin{figure}[htb]
 \vspace{-.4in}
 \hspace{.18in}\psfig{figure=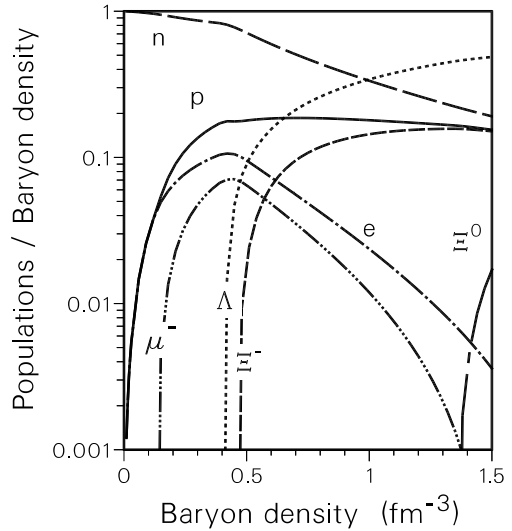,width=2.76in}
 \vspace{.2in}
 \begin{center}
 \parbox[t]{4.6 in} { \caption{ \label{popsk240sigma} Particle populations
 in neutron star matter containing nucleons, hyperons (absent the
  $\Sigma^-$ because of strong repulsion) and leptons.
  }}
  \end{center}
  \end{figure}

\section{RESULTS}

To illustrate that the behavior of the electrochemical potential is
only slightly influenced by the question of whether the $\Sigma^-$ hyperon 
experiences a strong repulsion in nuclear matter, we consider two
cases, in one of which all hyperons are coupled with the same strength
as the $\Lambda$, whose coupling can be constrained by observation as 
described above. In the other case, we consider the extreme case where the
$\Sigma^-$ experiences such a strong repulsion that it does not appear
at all in matter to densities exceeding those found in neutron stars.
To illustrate how hyperons arrest the growth of the electrochemical potential
with increasing density, we compare the above cases with a model in which
only nucleons and leptons appear. In the latter case, the electrochemical potential increases monotonically with density, and it is on that behavior that
the case for kaon condensation mainly rests. The results can be compared in 
Fig.\ \ref{mu3}.

It is apparent that the hyperons limit the growth of the  
electrochemical potential at a density of
2.5 to 3 times
nuclear density, and bring about its monotonic decrease at higher
density from a maximum value
of about 200 MeV, which
is far below the vacuum mass of the K$^-$ of 494 MeV.
This renders kaon condensation problematic, and further progress on this
question will require very accurate evaluation of the behavior of the
 K$^-$ mass as a function of density, as well as continuing experimental
 work on hyperon interactions.

\begin{figure}[htb]
\vspace{-.4in}
\hspace{.18in}\psfig{figure=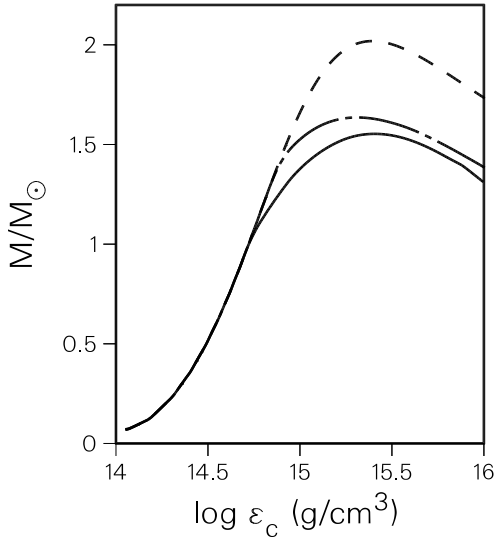,width=2.76in}
\vspace{.2in}
\begin{center}
\parbox[t]{4.6 in} { \caption{ \label{massigma} Neutron Star sequences 
corresponding to the three cases defined in Fig.\ \protect\ref{mu3}.
 }}
 \end{center}
 \end{figure}

It is interesting to see how the hyperon populations adjust to the possible
absence of the $\Sigma^-$. This can be seen by comparing Figs.\
\ref{pops_240} and \ref{popsk240sigma}. The second of these two figures
is the one in which the  $\Sigma^-$ is absent. We see that to compensate the
absence of the $\Sigma^-$, the $\Lambda$ threshold has been reduced somewhat,
and the $\Xi^-$ threshold has been greatly reduced. These changes take place to
most economically bring about charge neutrality in neutron star matter
and illustrate how powerful the Pauli Principle is in arranging
Fermion populations of conserved type in dense matter so as to minimize energy
at given density.
The great reduction of the $\Xi^-$ threshold in the absence of the  $\Sigma^-$
occurs because it is charge favored, replacing a neutron and electron at the
top of their Fermi seas (although both  $\Xi^-$ and $\Sigma^-$
are isospin unfavored) \cite{glen85:b}.
The threshold condition for baryon $B$ is
\begin{equation}
\mu_n \ge q_B \mu_e + g_{\omega B} \omega_0 +g_{\rho B} \rho_{03} I_{3B} +m_B
-g_{\sigma B} \,.
\end{equation}
The sign of $g_{\rho B} \rho_{03}$ is determined by the net isospin density
of the star, which is dominated by the neutron.
The first term on the left determines whether a given baryon charge state
is charge favored or unfavored, and the third term whether it is
isospin favored or unfavored.

The maximum neutron star mass is only somewhat perturbed by uncertainty in the 
$\Sigma^-$ coupling as can be seen in Fig.\ \ref{massigma}.
It is seen that hyperons significantly reduce the limiting neutron star
mass to a value $\sim 1.5 \msun$ in this theory with coupling constants
chosen in accord with nuclear and hypernuclear data.  The latter data
is not nearly as firm as the former and introduces some uncertainty. 
A limit of $\sim 1.7 \msun$ for neutron stars would be compatible
with these uncertainties, but is in our estimation less favored than
the first limit mentioned \cite{glen91:c}.\\[2ex]

\doe



\begin{thebibliography}{10}

\bibitem{glen91:c}
N. K. Glendenning and S. A. Moszkowski, Phys.\ Rev.\ Lett.\ {\bf 67} (1991)
  2414.

\bibitem{dover88:a}
D. J. Millener, C. B. Dover and A. Gal, Phys.\ Rev.\ C {\bf 38} (1988) 2700.

\bibitem{rufa90:a}
M. Rufa, J. Schaffner, J. Marhun, H. Stocker, W. Greiner and P.-G. Reinhard,
  Phys.\ Rev.\ C {\bf 42} (1990) 2469.

\bibitem{glen85:b}
N. K. Glendenning, Astrophys.\ J.\ {\bf 293} (1985) 470.

\bibitem{kaplan86:a}
D. B. Kaplan and A. Nelson, Phys.\ Lett.\ {\bf 175 B} (1986) 57.

\bibitem{schaffner96:a}
J. Schaffner and I. N. Mishustin, Phys.\ Rev.\ C {\bf 53} (1996) 1416.

\bibitem{schaffner96:b}
J. Schaffner, J, Bondorf and I. N. Mishusten, Heavy Ion Physics, {\bf 4} (1996)
  293.

\bibitem{balberg97:a}
S. Balberg and A. Gal, Nucl. Phys {\bf A 625} (1997) 435.

\bibitem{gal94:a}
E. Friedmann, A. Gal and C. J. Batty, Nucl.\ Phys.\ {\bf A 579} (1994) 518.

\bibitem{brown}
G. E. Brown, Private communication, 1999.

\bibitem{dover83:a}
C. B. Dover and A. Gal, Ann.\ Phys.\ (N.Y.) {\bf 146} (1983) 309.

\bibitem{fukuda}
T. Fukuda {\sl et. al.} Phys.\ Rev.\ C {\bf 58} (1998) 1306.

\bibitem{khaustov}
P. Khaustov {\sl et. al.} Phys.\ Rev.\ C {\bf 61} (2000) 054603.

\bibitem{books}
N.\ K.\ Glendenning, {\sl COMPACT STARS} (Springer--Verlag New York, 1'st ed.
  1997, 2'nd ed. 2000).

\bibitem{glen82:a}
N. K. Glendenning, Phys.\ Lett.\ {\bf 114B} (1982) 392.

\bibitem{bodmer77:a}
J. Boguta and A. R. Bodmer, Nucl.\ Phys.\ {\bf A292} (1977) 413.

\bibitem{serot87:a}
B. D. Serot and H. Uechi, Ann. Phys. (New York) {\bf 179} (1987) 272.

\bibitem{kapusta90:a}
J. I. Kapusta and K. A. Olive, Phys.\ Rev.\ Lett.\ {\bf 64} (1990) 13.

\bibitem{kapusta91:a}
J. Ellis, J. I. Kapusta and K. A. Olive, Nucl.\ Phys.\ {\bf B348} (1991) 345.

\bibitem{teller55:a}
M. H. Johnson and E. Teller, Phys.\ Rev.\ {\bf 98} (1955) 783.

\bibitem{duerr56:a}
H. P. Duerr, Phys.\ Rev.\ {\bf 103} (1956) 469;.

\bibitem{walecka74:a}
J. D. Walecka, Ann.\ of Phys.\ {\bf 83} (1974) 491.

\end{thebibliography}

\end{document}